 \documentclass[prd,preprintnumbers,amsmath,amssymb,floatfix]{revtex4}

\usepackage{graphicx}
\usepackage{dcolumn}
\usepackage{bm}
\usepackage{amssymb}
\usepackage{epsfig}
\usepackage{color}

\newcommand{\ba}{\begin{eqnarray}}
\newcommand{\ea}{\end{eqnarray}}
\newcommand{\be}{\begin{equation}}
\newcommand{\ee}{\end{equation}}
\newcommand{\bdisplay}{\begin{displaymath}}
\newcommand{\edisplay}{\end{displaymath}}

\newcommand{\eq}[1]{Eq.\,(\ref{#1})}

\newcommand{\calF}{\mbox{${\cal F}$}}

\newcommand{\delchi}{\Delta \chi^2_i}

\newcommand{\delchimax}{{\delchi}_{\rm max}}

\newcommand{\Fp}{$F_2^{\gamma p}(x,Q^2)$}

\newcommand{\hatF}{\mbox{${\hat {\cal F}}(v,Q^2)$}}

\begin{document}

\title{  A new numerical method for obtaining gluon distribution functions $G(x,Q^2)=xg(x,Q^2)$, from the proton structure function $F_2^{\gamma p}(x,Q^2)$.}  

\author{Martin~M.~Block}
\affiliation{Department of Physics and Astronomy, Northwestern University, 
Evanston, IL 60208}

\date{\today}

\begin{abstract}
An exact expression for the leading-order (LO) gluon distribution function $G(x,Q^2)=xg(x,Q^2)$ from the DGLAP evolution equation for the proton structure function $F_2^{\gamma p}(x,Q^2)$ for deep inelastic $\gamma^* p$ scattering has recently been obtained [M. M. Block, L. Durand and D. W. McKay, Phys. Rev. D{\bf 79}, 014031, (2009)] for massless quarks, using  Laplace transformation techniques.  Here, we develop a fast and accurate numerical inverse Laplace transformation algorithm, required  to invert the Laplace transforms needed to evaluate $G(x,Q^2)$, and compare it to the exact solution.  We obtain accuracies of less than 1 part in 1000 over the entire $x$ and $Q^2$ spectrum.  Since no analytic Laplace inversion is possible for next-to-leading order (NLO) and higher orders, this numerical algorithm will enable one to obtain  accurate  NLO (and NNLO) gluon distributions, using only experimental measurements  of $F_2^{\gamma p}(x,Q^2)$.
\end{abstract}

\maketitle


\section{Introduction} \label{sec:introduction} 

The quark and gluon distributions in hadrons  play a key role in our understanding of Standard Model processes, in our predictions for such processes at accelerators, and in our searches for new physics.  In particular, accurate knowledge of gluon distribution functions at small Bjorken $x$ will play a vital role in estimating backgrounds, and hence, our ability to search for new physics at the Large Hadron Collider.  

The gluon and quark distribution functions have traditionally been determined  simultaneously by fitting experimental data  (mainly at small $x$) on the  proton structure function $F_2^{\gamma p}(x,Q^2)$ measured in deep inelastic $ep$ (or $\gamma^*p$) scattering, over a large domain  of values of $x$ and $Q^2$.  The process starts with an initial  $Q^2_0$, typically in the 1 to 2 GeV$^2$ range, and  individual quark and gluon trial distributions parameterized  as functions of $x$. The distributions are evolved to larger $Q^2$ using the coupled integral-differential DGLAP equations  \cite{dglap1,dglap2,dglap3}, and the results used to predict the measured quantities. The final distributions are then determined by adjusting the input parameters to obtain a best fit to the data. For recent determinations of the gluon and quark distributions, see \cite{CTEQ6.1,CTEQ6.5,MRST,MRST4}.  

This procedure is rather indirect, especially so in the case of the gluon: the gluon distribution $G(x,Q^2)$ does not appear in the experimentally accessible quantity $F_2^{\gamma p}(x,Q^2)$, and is determined only through the quark distributions in conjunction with the evolution equations.  It is  further not clear without detailed analysis \cite{MRST3,CTEQchi2,CTEQ_Hessian,CTEQ6.5} how sensitive the results are to the parameterizations of the initial parton distributions, or how well the gluon distribution is actually determined.

In a recent paper,  Block, Durand and McKay (BDM) \cite{bdm2}  used a Laplace transformation technique to obtain a  leading order (LO) analytic gluon distribution function $G(x,Q^2)=xg(x,Q^2)$  for massless quarks, {\em directly} from a global parameterization of the data on  $F_2^{\gamma p}(x,Q^2)$.   The method uses only the LO DGLAP evolution equation \cite{dglap1,dglap2,dglap3} for $F_2^{\gamma p}(x,Q^2)$ in the usual approximation in which the active quarks are treated as massless. In contrast to previous methods for determining $G(x,Q^2)$, it  does not require knowledge of the separate quark distributions in the region in which structure function data exist, nor does it require the use of the evolution equation for $G(x,Q^2)$, both considerable simplifications. In essence, the authors transform the LO DGLAP equation from Bjorken $x$-space to $v$-space, where $v\equiv \ln(1/x)$,  and then Laplace transform the resulting equation. After solving for the Laplace transform $g(s,Q^2)$, they are able to  analytically invert the Laplace transform back into $v$-space, and eventually, to $x$-space. Unfortunately, because of the considerable complexities of next-to-leading order (NLO) splitting functions needed in the NLO DGLAP evolution equation \cite{dglap1,dglap2,dglap3} for $F_2^{\gamma p}(x,Q^2)$ , the method can {\em not} be extended to NLO gluon distributions, because of the impossibility of analytically inverting the required Laplace transform.  

The purpose of this note is to derive a new and fast algorithm for accurate numerical inversion of Laplace transforms, so that the BDM method  \cite{bdm2} for direct evaluation of gluon distributions from  knowledge of $F_2^{\gamma p}(x,Q^2)$  can be extended to NLO (and NNLO) easily and accurately. Mathematically, Laplace inversion is an ``ill-posed'' problem and great care must be taken to insure its reliability for arbitrary Laplace transforms \cite{Graf}.   In this case, we can check our numerical Laplace inversion routine directly by comparing its results to the exact LO gluon solution of Ref. \cite{bdm2}.

\section{The exact LO solution} \label{sec:LOsolution} 
The LO DGLAP equation for the evolution of the proton structure function \Fp\   for 4 massless quarks ($u,d,s,c$) can be written as
\be
\label{evolution}
\frac{\partial F_2^{\gamma p}(x,Q^2)}{\partial\ln Q^2}  = \frac{\alpha_s}{4\pi}\left[x \int_{x}^1\frac{dz}{z^2}F_2^{\gamma p}(z,Q^2)K_{qq}\left(\frac{x}{z}\right) +{20\over 9} x\int_{x}^1\frac{dz}{z^2}G(z,Q^2)K_{qg}\left(\frac{x}{z}\right)\right],
\ee
where $K_{qq}(x)$ and $K_{qg}(x)$ are the LO splitting functions and $\alpha_s$ is the strong running coupling constant.

Following BDM \cite{bdm2}, we introduce ${\cal F}_2^{\gamma p}(x,Q^2)$
\ba
{\cal F}_2^{\gamma p}(x,Q^2) &\equiv& \frac{\partial F_2^{\gamma p}(x,Q^2)}{\partial\ln Q^2}-{\alpha_s\over 4 \pi}x \int_{x}^1\frac{dz}{z^2}F_2^{\gamma p}(z,Q^2)K_{qq}\left(\frac{x}{z}\right).
 \label {LOAP2} 
\ea

We finally write the DGLAP equation for the evolution of $F_2^{\gamma p}(x,Q^2)$ as 
\ba
x\int_x^1G(z,Q^2)K_{qg}\left({x\over z}\right)\frac{\,dz}{z^2}&=&{\cal F}(x,Q^2), \label{Geqn}
\ea
where
\be
{\cal F}(x,Q^2)\equiv {9\over 20}\left ({ \alpha_s\over 4\pi}\right)^{-1}{\cal F}_2^{\gamma p}(x,Q^2).\label{FF}
\ee
and the LO $g\rightarrow q$ splitting function is given by
\ba
K_{qg}(x)=1-2x+2x^2. \label{splittingfunction}
\ea
At this point,  BDM introduce the coordinate transformation 
\be v\equiv \ln(1/x),
\ee
and  define functions $\hat{G}$, $\hat{K}_{qg}$, and $\hat{\cal F}$ in $v$-space by
\ba
\hat G(v,Q^2)&\equiv& G(e^{-v},Q^2)\nonumber\\
\hat K_{qg}(v)&\equiv& K_{qg}(e^{-v})\nonumber\\
\hat \calF (v,Q^2)&\equiv& {\calF}(e^{-v},Q^2).\label{hats}
\ea
Explicitly, from \eq{splittingfunction}, we see that
\ba
\hat K_{qg}(v)=1-2e^{-v}+2e^{-2v}\label{Khat}. 
\ea
Since
\ba
\hatF\!&=&\!\int^v_0\hat G(w,Q^2)e^{-(v-w)}\hat K_{qg}(v-w)\,dw,
\ea
 BDM note that in $v$-space, the DGLAP equation of \eq{Geqn}  can be written as the convolution integral
\ba
\hatF\!&=&\! \int^v_0\hat G(w,Q^2)\hat H(v-w)\, dw,\label{DGLAP1}
\ea
where
\ba
\hat H(v)&\equiv&e^{-v}\hat K_{qg}(v)\nonumber\\
&=& e^{-v}-2e^{-2v}+2e^{-3v}.\label{H}
\ea

The  Laplace transform of  $\hat H(v)$ is given by $h(s)$, where 
\ba
h(s)\equiv{\cal L}[\hat H(v);s]= \int_0^\infty \hat H(v)e^{-sv} dv,
\quad {\rm with\ }\hat H(v)=0,\   v<0.\label{Laplacetransform}
\ea
 
The convolution theorem for Laplace transforms allows us to rewrite the right hand side of \eq{DGLAP1} as a product of their Laplace transforms $g(s)$ and $h(s)$, so that the Laplace transform of \eq{DGLAP1} is given by the algebraic equation 
\ba
f(s,Q^2)&=&g(s,Q^2)\times h(s).\label{LT1}
\ea
  Solving \eq{LT1} for $g$, the Laplace transform of the gluon distribution function in $s$-space is given by
\be
g(s,Q^2)=f(s,Q^2)/h(s). \label{g_solution}
\ee
In general, one is not able to calculate the inverse transform of $g(s,Q^2)$ explicitly, if only because $f(s,Q^2)$ is determined by a numerical integral of the experimentally-determined function $\hat{\cal F}(v,Q^2)$. However, regarding $g(s,Q^2)$ as the product of the two functions $f(s,Q^2)$and $h^{-1}(s)$ and taking the inverse Laplace transform using the inverse of the convolution theorem and the {\em known inverse} ${\cal L}^{-1}[f(s,Q^2);v]={\hat \calF}(v,Q^2)$, BDM find the analytic solution
\ba
{\hat G}(v,Q^2)&=&{\cal L}^{-1}[f(s,Q^2)\times h^{-1}(s);v].
\ea

The calculation of $h(s)$ and the inverse Laplace transform of $h^{-1}(s)$ are straightforward, and the answer for LO, albeit singular, is given by BDM in terms of the Dirac delta function as
\ba
{\cal L}^{-1}[h^{-1}(s);v]&=&3\delta(v)+ \delta'(v)-e^{-3 v/2}\left({6\over \sqrt 7} \sin \left[ {\sqrt 7\over 2}v\right]\nonumber +2\cos\left[{\sqrt 7\over 2}v\right] \right).\label{Hminus1}
\ea
Thus, again using the convolution theorem, BDM find that  the analytic  solution for the LO gluon distribution $\hat G(v,Q^2)$ for massless quarks  is
\ba
\hat G(v,Q^2)&= &3{\hat \calF}(v,Q^2)+{\partial {\hat \calF}(v,Q^2)\over \partial v }\nonumber\\
&&\quad-\int_0^v {\hat\calF}(w,Q^2) e^{-3 (v-w)/2}\times\left({6\over \sqrt 7} \sin \left[ {\sqrt 7\over 2} (v-w)\right]+2\cos\left[{\sqrt 7\over 2}(v-w)\right]\! \right)dw.\label{Gofv}
\ea
The BDM  solution depended critically upon the ability to find the analytic inverse Laplace transform of $h^{-1}(s)$, which was possible for the LO case of the splitting function $K_{qg}(x)$. 

Unfortunately, for NLO or higher, the splitting function is so complicated that an analytic inversion for the NLO  $h^{-1}$ is impossible---see Floratos et al. \cite{floratos} for the NLO $ {\rm \overline {MS}}$ splitting function that is required.  For this case, we must be able to find a {\em numerical} inversion of the Laplace transform $g(s)$ of the equivalent of \eq{g_solution} in order to obtain $\hat G(v,Q^2)$ and, ultimately, $G(x,Q^2)$. The goal of this communication is to develop a suitable  numerical Laplace inversion algorithm.
\section{Numerical inversion of Laplace transforms}\label{sec:inverseLaplace}

In order to simplify our notation, we will now suppress the explicit dependence of $\hat G(v,Q^2)$ on $Q^2$, writing it as $\hat G(v)$. If the Laplace transform of $g(s)\equiv {\cal L}[\hat G(v);s]=\int_0^\infty G(v)e^{-vs}\,dv$, where $\hat G(v)=0$ for $v<0$,  then the inverse Laplace transform is given by the complex Bromwich integral
\ba
\hat G(v)\equiv {\cal L}^{-1}[g(s);v]={1\over 2\pi i} \int^{\,c+i\,\infty}_{\,c-i\,\infty}g(s)e^{vs}\,ds,\label{Bromwich}
\ea
 where the real constant $c$ is to the right of all singularities of $g(s)$. We will further assume that we have made an appropriate coordinate  translation in $s$ so that $c=0$, so that the equation is written as
\ba
\hat G(v)\equiv {\cal L}^{-1}[g(s);v]={1\over 2\pi i} \int^{\,+i\,\infty}_{\,-i\,\infty}g(s)e^{vs}\,ds.\label{Bromwich2}
\ea
 Our goal is to {\em numerically} solve \eq{Bromwich2}. The inverse Laplace transform is essentially determined by the behavior of $g(s)$  near its  singularities, and thus is an ill-conditioned or ill-posed numerical problem. We suggest in this note a new algorithm that takes advantage of very fast, arbitrarily  high precision complex number arithmetic that is possible today in programs like {\em Mathematica} \cite{Mathematica}, making the inversion problem numerically tractable. 

First, we introduce a new complex variable $z\equiv vs$ and rewrite \eq{Bromwich2} as
\ba
\hat G(v)&=&{1\over 2\pi i v} \int^{\,+i\,\infty}_{\,-i\,\infty}g\left({z\over v}\right)e^{z}\,dz.\label{Bromwichtransformed}
\ea

We next make a rational approximation to  $e^{z}$,  using  the partial fraction expansion
\ba
e^z\approx \sum_{i=1}^{2N}{\omega_i\over z- \alpha_i}\label{exp_expansion},
\ea
which can be shown to have the following properties:
\begin{enumerate}
  \item $\rm{Re}\ \alpha_1>0$, so that its poles are all in the right-hand half of the complex plane. 
  \item $N$ distinct complex conjugate pairs of the complex numbers $(\omega_i,\alpha_i)$, such that the sum of the $k^{\rm th}$ pair,
 \be{\omega_k\over z- \alpha_i}+{\bar \omega_k\over z- \bar\alpha_i},
\ee
 is real for all real $z$. 
\item The expansion is identical to the Pad\'e approximant with numerator equal to $2N-1$ and denominator equal to $2N$. 
 \item The integrand vanishes {\em faster} than $1/R$ as $R\rightarrow\infty$ on the   semi-circle of radius $R$ that encloses the right hand half of the complex plane, since the approximation vanishes as $1/R$ and $g(s)$ that corresponds to a non-singular $G(v)$ also vanishes for $R\rightarrow\infty$.  
\end{enumerate}
Since g(z/v)  must vanish for $z\rightarrow\infty$  and our approximation      for $e^z$ in \eq{exp_expansion} {\em vanishes} for $z\rightarrow\infty$, we can form a  closed contour $C$ by completing  our integration path of the modified  Bromwich integral in \eq{Bromwichtransformed} with an infinite half circle to the {\em right} half of the complex plane. As mentioned earlier, $g(z/v)$ has no singularities in this half of the complex plane. It is important to note that this contour is  a {\em clockwise} path  around the poles of \eq{exp_expansion}, which come  from  our approximation to $e^z$. What we need is the negative of it, i.e., the contour $-C$ which is counterclockwise, so that the poles are to our left as we traverse the contour $-C$. 
Therefore, we rewrite \eq{Bromwichtransformed} as 
\ba
\hat G(v)&\approx& {1\over 2\pi i v} \oint_C g\left({z\over v}\right)\sum_{i=1}^{2N}{\omega_i\over z-\alpha_i}\,dz\nonumber\\
&=&-{1\over 2\pi i v} \sum_{i=1}^{2N}\oint_{-C} g\left({z\over v}\right){\omega_i\over z-\alpha_i}\,dz\nonumber\\
&= &-{2\over  v} \sum^{N}_{i=1}{\rm Re}\left[\omega_ig\left({\alpha_i/ v}\right)\right].\label{Bromwich3}
\ea
To obtain \eq{Bromwich3}, the final approximation to $\hat G(v)$,   we used Cauchy's theorem  to equate the closed contour integral around the path $-C$ to $2\pi i$ times the sum of the (complex) residues of the poles.  Since the contour $-C$ restricts us to the right-hand half of the complex plane, no poles of $g(z/v)$ were enclosed, but only the $2N$ poles $\alpha_i$ of the approximation of $e^z$. Using the properties cited above of the  complex conjugate pairs---$(\omega_i,\alpha_i)$ and  $(\bar \omega_i ,\bar \alpha_i)$---after taking only their real part and multiplying by 2,  we have simultaneously insured that $\hat G(v)$ is real , yet only have had to sum over half of the residues.

Equation (\ref{Bromwich3}) has some very interesting properties:

\begin{enumerate} 
  \item The $4N$ coefficients ($\alpha_i,\omega_i)$ are  complex constants that are  independent of $v$, only depending on the value of $2N$ used for the approximation, so that for a given $2N$, they only have to be evaluated once---in essence, they can be tabulated and stored for later use.  
  \item The Laplace transform of $v^n$ is given by ${n!/s^{n+1}}$, where $n$ is integer. Inserting $G(v)=v^n$ and $g(s)={n!/ s^{n+1)}}$ into \eq{Bromwich3}, we see that we have a set of $4N$ equations, 
\ba
-n!\sum_i^N 2 {\rm Re}\left({\omega_i\over \alpha_i^{n+1}} \right)=1,\quad n=0,1,\ldots ,4N-1,\label{simultaneous}
\ea
which also uniquely determine the $4N$ complex constants $(\alpha_i,\omega_i),\quad i=1,2,\ldots,2N$, although evaluating them directly from  \eq{simultaneous} is virtually impossible numerically, considering the ill-posed nature of these equations.

The true power of \eq{simultaneous} that it shows that  the inversion algorithm of \eq{Bromwich3} is {\em exact} when $G(v)$ is a polynomial of order $\le 4N-1$, even though only $N$ complex terms have to be evaluated  in \eq{Bromwich3}. This is reminiscent of the situation using  Gauss-Legendre integration of order $N$, where there are $2N$ constants, $N$ Legendre zeroes and N weights, and the  integration  approximation is {\em exact} if the integrand is a polynomial of order $\le 2N-1$.  
\item 
The real parts of the residues  $\omega_i$  alternate in sign and are {\em exceedingly} large---even for relatively modest $2N$, making round-off a potentially serious problem.  Thus, exceedingly high precision complex arithmetic is called for, often requiring  60 or more digits.  However, this is not a serious problem---either in speed or complexity of execution---for an algorithm written in a program such as {\em Mathematica} \cite{Mathematica}.
\end{enumerate}

A concise inversion algorithm in {\em  Mathematica} that implements \eq{Bromwich3} is  given in Appendix \ref{subsection:inversionalgorithm} and a one line {\em Mathematica} algorithm for finding a numerical Laplace transform is given in Appendix \ref{subsection:Laplacetransform}.

We will now test  the accuracy of our numerical Laplace inversion algorithm by comparing its results with \eq{Gofv}, the exact  LO $G(v,Q^2)$ of BDM.  


\section{Comparison of exact solution and numerical Laplace inversion results}

\begin{figure}[h,t,b] 
\begin{center}
\mbox{\epsfig{file=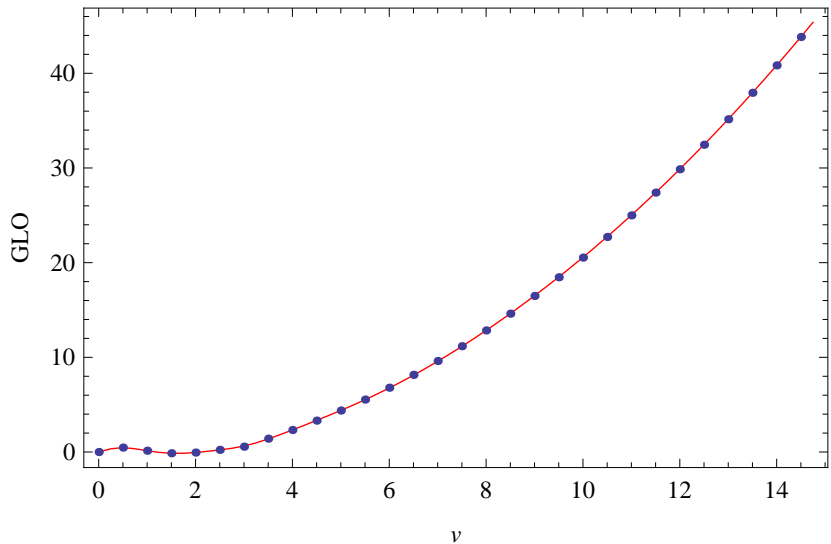
,width=6in%
,bbllx=0pt,bblly=0pt,bburx=420pt,bbury=198pt,clip=%
}}
\end{center}
\caption[]{LO gluon distribution functions $\hat G(v,Q^2)$ vs. $v =\ln (1/x)$, for virtuality $Q^2=  5$ GeV$^2$. The blue points are the exact LO solution of \eq{Gofv}. The red curve is our {\em numerical}   Laplace  inversion solution.  The agreement is excellent over the entire $v$ range, which corresponds to the $x$-range, $5\times 10^{-7}\le x\le 1$.
\label{fig:GLO5}}
\end{figure}

\begin{figure}[h,t,b] 
\begin{center}
\mbox{\epsfig{file=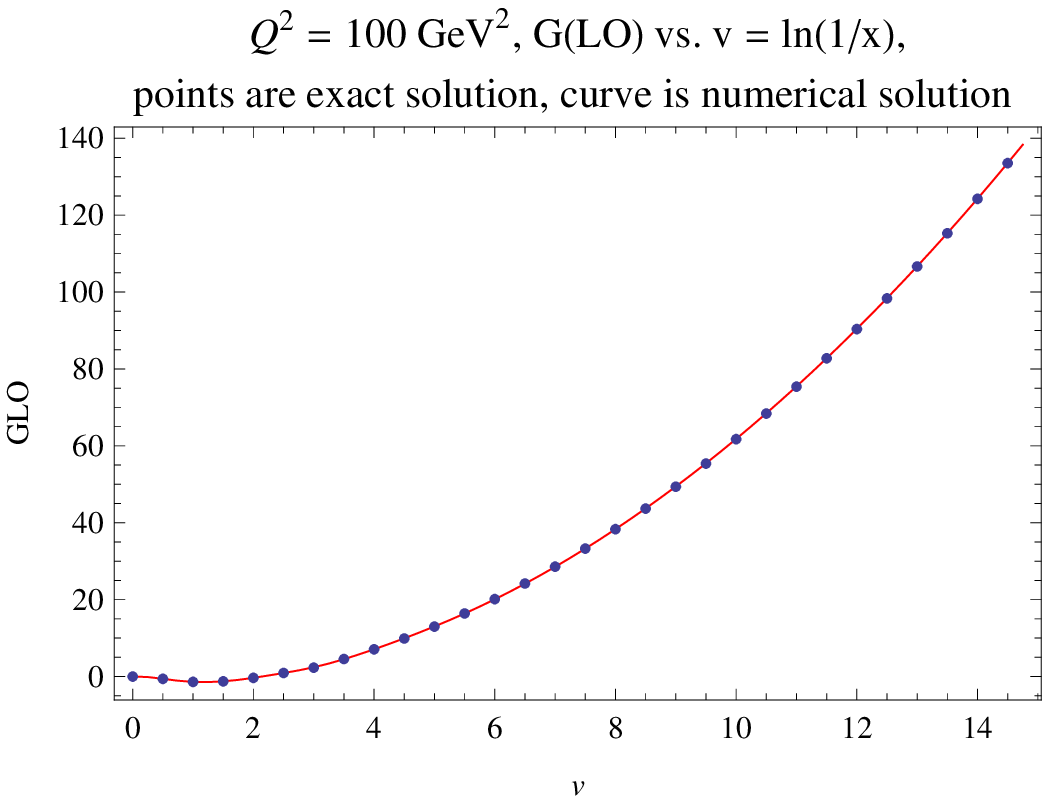
,width=6in%
,bbllx=0pt,bblly=0pt,bburx=420pt,bbury=198pt,clip=%
}}
\end{center}
\caption[]{LO gluon distribution functions $\hat G(v,Q^2)$ vs. $v =\ln (1/x)$, for virtuality $Q^2=  100$ GeV$^2$. The blue points are the exact LO solution of \eq{Gofv}. The red curve is our {\em numerical}   Laplace  inversion solution.  The agreement is excellent over the entire $v$ range, which corresponds to the $x$-range, $5\times 10^{-7}\le x\le 1$.
\label{fig:GLO100}}
\end{figure}

Using the Berger, Block and Tan \cite{bbt2}  fit to the experimental ZEUS 
\cite{ZEUS2} data shown in Appendix \ref{subsection:ZEUS}, we have calculated both the exact solution for $\hat G(v,Q^2)$ from \eq{Gofv} and the completely numerical approximation  for $\hat G(v)$  given in  \eq{Bromwich3},  using our inverse Laplace transformation  algorithm given in Appendix \ref{subsection:inversionalgorithm}.  For this purpose, we used {\texttt 2N} = 8 and {\texttt prec} = 80 in the algorithm.  The results for $Q^2=5$ GeV$^2$ and 100 GeV$^2$ are shown in Fig. \ref{fig:GLO5} and Fig. \ref{fig:GLO100}, respectively. The blue points are the exact solution and the red curves are the numerical solution that uses our algorithm for the numerical inversion of Laplace transforms.  As seen in both Figures, the agreement is striking over the entire $v$ range, which corresponds to the $x$ interval  $5\times 10^{-7}\le x\le 1$.  At the highest $v$ values (where numerical Laplace inversion approximations generally  have the greatest problem), we find an accuracy of $\sim$ 1 part in 5000, for $Q^2=100$ GeV$^2$. This error reflects both the numerical inaccuracy associated with the numerical integration term in the exact solution  $\hat G(v,Q^2)$  of \eq{Gofv}, as well as the inaccuracies associated with the numerical Laplace inversion routine of Appendix \ref{subsection:inversionalgorithm} and the numerical Laplace transformation routine of Appendix \ref{subsection:Laplacetransform}.  

Another independent method of checking the numerical accuracv of the entire  procedure is to go back to the original  DGLAP equation from which we started, \eq{Geqn}, i.e.,
\ba
x\int_x^1G(z,Q^2)K_{qg}\left({x\over z}\right)\frac{\,dz}{z^2}&=&{\cal F}(x,Q^2), \label{Geqn2}
\ea
and numerically integrate  its l.h.s., which depends on our numerical solution for $\hat G(v,Q^2)$ through 
$G(x,Q^2)=\hat G(\ln(1/x),Q^2)$. We then  compare  it with the r.h.s., ${\cal F}(x,Q^2)$, which is independently known for all $x$ and $Q^2$.   This check becomes of primary importance when one does {\em not} have an analytical solution---the typical situation.  In this case, it validated our conclusions about the numerical accuracy of our inversion procedure over the entire $x$ and $Q^2$ domain.  In particular, for $Q^2=100$ GeV$^2$, the ratio of the l.h.s. to the r.h.s. of \eq{Geqn2} is unity to 1 part in 5000 in the x range $3\times 10^{-4}< x < 3\times 10^{-2}$, and never rises to more than $\sim$ 1 part in 200 outside this range;  these excursions from unity include the errors generated in the  numerical integration of the l.h.s of \eq{Geqn2}. 

In contrast to the highly singular behavior of  the analytic Laplace inverse of inversion of $h^{-1}(s)$ in \eq{Hminus1}, the overall behavior of $\hat G(v,Q^2)$, the Laplace inversion of $f(s,Q^2)/h(s)$  in \eq{Gofv},  is very smooth and therefore lends itself readily to numerical inversion, as seen by the excellent agreement  shown in Fig. \ref{fig:GLO5} and Fig. \ref{fig:GLO100} between the exact solutions and the numerical inverse Laplace transforms. 
 
Although our inversion routine was specifically developed to invert Laplace transforms of gluon distributions, it clearly has a wide variety of applications in solving both integral and differential equations, which will not be commented on further in this communication.


\section{Conclusions}
We have achieved high numerical accuracy in obtaining LO gluon distributions from fits to experimental data, using a numerical Laplace inversion routine developed for this purpose, and have compared it successfully to the exact analytic solution using the same fit.   Since NLO and NNLO DGLAP solutions are not amenable to having analytic solutions---one can {\em not} analytically invert their $h^{-1}(s)$---this technique will allow us to find very accurate  numerical  gluon solutions {\em directly} from experimental data, without having first  to find quark distributions by  solving coupled DGLAP equations.  Further, we will be able to verify the numerical accuracy of our solutions by direct substitution into their generating equations.  
\appendix 
\section{{\em Mathematica} Laplace  Inversion Algorithm}\label{subsection:inversionalgorithm}
\texttt{
NInverseLaplaceTransformBlock[g\_,s\_,v\_,twoN\_,prec\_]:=Module[\\
\hphantom{xxxxx}\{Omega,Alpha,M,p,den,r,num\},\\
\hphantom{xxxxx}(M=2*Ceiling[twoN/2];p=PadeApproximant[Exp[z],\{z,0,\{M-1,M\}\}];\\
\hphantom{xxxxx}den=Denominator[p];r=Roots[den==0,z];
Alpha=Table[r[[i,2]],\{i,1,M\}];\\
\hphantom{xxxxx}num=Numerator[p];hospital=num/D[den,z];\\
\hphantom{xxxxx}Omega=SetPrecision[Table[hospital/.z->Alpha[[i]],\{i,1,M,2\}],prec+50];\\
 \hphantom{xxxxx}Alpha=SetPrecision[Table[Alpha[[i]],\{i,1,M,2\}],prec]);\\
\hphantom{xxxxx}SetPrecision[-(2/v)Sum[Re[Omega[[i]] g/.s->Alpha[[i]]/v],\{i,1,M/2\}],30]
\\
\hphantom{xxxxxxxxxxxxxxxxxxxxxxxxxxxxxxxxxxxxxxXXXXxxxxxxxxxxxxxxxxxxx}]}

In the above algorithm, \texttt{g} = $g(s)$, \texttt{s} = $s$, \texttt{v} = $v$, \texttt{twoN}=$2N$ in \eq{Bromwich3}, and \texttt{prec} = precision of calculation. Typical values are \texttt{twoN}  = 8 and 
\texttt{prec} = 70.  The algorithm, which is quite fast, returns the numerical value of $\hat G(v)$. It utilizes that fact that the partial fraction approximation made for $e^z$ is equal to a Pad\'e approximant whose numerator is order $2N-1$ and whose denominator is order $2N$.  

The algorithm first insures that  \texttt{M=twoN} is {\em even}. It then constructs \texttt{p}, the Pad\'e approximant whose numerator is a polynomial of order $M-1$ and denominator a polynomial of order $M$. It finds \texttt{r}, the complex roots of the denominator, which are $\alpha_i$, the poles  of \eq{Bromwich3}.  Using L'Hospital's rule, it finds the residue $\omega_i$ corresponding to the pole $\alpha_i$. At this point, all of the mathematics  is symbolic. It next finds every {\em other} pair of $(\alpha_i,\omega_i)$ to the desired numerical accuracy; they come consecutively, i.e., $\alpha_1=\bar \alpha_2,\ \omega_1 = \bar \omega_2,\ \alpha_3=\bar \alpha_4,\ \omega_3 = \bar \omega_4$, etc.  Finally, it  takes the necessary sums, again to the desired numerical accuracy, but only  over  half of the interval $i=1,3,\ldots,N$, by taking  only the real part and multiplying by 2. 

If $g(s)$, the input to the algorithm, is an {\em analytic} relation {\em and} $v$ is a pure number (from the point of view of {\em Mathematica}, 31/10 is a pure number, but 3.1 is {\em not}), then, for sufficiently high values of {\tt prec}, you can achieve arbitrarily high accuracy. If we define the accuracy as $1-G(v)_{\rm numerical}/G(v)_{\rm true}$, numerical tests on many different functions shows that it goes to 0 for large {\tt 2N} as an inverse power law in {\tt 2N}. On the other hand, if $g(s)$ is obtained {\em numerically}, either from having to find the Laplace transform $f(s)$ and/or $h(s)$ from {\em numerical integration} techniques, the overall accuracy of inversion is limited by the need to only use relatively small values of {\tt 2N}---in the neighborhood of $6-12$, limiting the overall accuracy to be in the neighborhood of $10^{-5}$, which fortunately is ample for most numerical work. Typically, numerical integration routines are not accurate to better than $\sim 10^{-5}$, and one can not use  $\omega$'s---which alternate in sign---that are larger than $\sim 10^{14}-10^{16}$, which occur for relatively small values of {\tt 2N}. Of course, this is {\em not} a limitation if $g(s)$ is able to be expressed in closed form.

\section{{\em Mathematica} Numerical Laplace Transform Algorithm}\label{subsection:Laplacetransform}
We note from \eq{g_solution} that the function that we must invert is
\be
g(s)=f(s)/h(s). \label{g_solution2}
\ee
Although we can obtain an analytic solution for $h(s)$---true for LO as well as NLO---we must find $f(s)$ by numerical means.
From \eq{Laplacetransform}, we see that the Laplace transform of $\hat F(v)$ is given by the integral 
\ba
f(s)= \int_0^\infty \hat F(v)e^{-sv} dv.\label{Laplacetransform2}
\ea
Because of the $\infty$ at the upper limit of the integral, for numerical work it is useful to make the transformation $v=-\ln u$, yielding
\ba
h(s)= \int_0^1 \hat H(-\ln u)u^{s-1} du.\label{Laplacetransform3}
\ea
The upper limit of the integral is now finite,   
leading to the stable {\em Mathematica} algorithm:

\texttt{NLaplaceTransform[H\_, vv\_, ss\_] := 
 NIntegrate[u\^}$\mathtt{\,}$\texttt{ (ss - 1)*H /.vv -> -Log[u], \{u, 0, 1\}, 
  AccuracyGoal -> 15, MaxRecursion -> 15]}

 \section{Global parameterization of $F_2^{\gamma p}(x,Q^2)$ using ZEUS structure function data}\label{subsection:ZEUS}
Berger, Block and Tan \cite{bbt2} have parameterized the proton structure function $F_2^{\gamma p}(x,Q^2)$ as
\ba
F_2^{\gamma p}(x,Q^2)=(1-x)&& \hspace*{-1em} \Bigg\{\frac{F_P}{1-x_P}+A(Q^2)\ln\left[\frac {x_P}{x}\frac{1-x}{1-x_P}\right] \nonumber \\
& + & 
B(Q^2)\ln^2\left[\frac {x_P}{x}\frac{1-x}{1-x_P}\right]\Bigg\}. \label{Fofx} 
\ea
Here $x_P=0.09$ specifies the location in $x$ of an approximate fixed point observed in the data where curves for different $Q^2$ cross. At that point, $\partial F_2^{\gamma p}(x_P,Q^2)/\partial \ln Q^2\approx 0$ for all $Q^2$;  $F_P=F_2^{\gamma p}(x_P,Q^2)=0.41$ is the common value of $F_2^{\gamma p}$.  The $Q^2$ dependence of $F_2^{\gamma p}(x,Q^2)$ is  given in those fits by
\ba 
    A(Q^2)&=&a_0+a_1\ln Q^2 +a_2\ln^2 Q^2, \nonumber \\ 
    B(Q^2)&=&b_0+b_1\,\ln Q^2 +b_2\,\ln^2 Q^2.  \label{AB}
\ea

The fitted quantities and their errors are shown in Table \ref{fitted}.
\begin{table}[h]                   
%
\begin{center}
\def\arraystretch{1.2}            
     \caption{\label{fitted}\protect\small Results of a 6-parameter fit to ZEUS $F_2^p(x,Q^2)$ structure function data \cite{ZEUS2} using the $x$ 
and $Q^2$ behaviors of \eq{Fofx} and \eq{AB}, with $Q^2$ in GeV$^2$.   The renormalized 
$\chi^2_{\rm min}$ per degree of freedom, taking into account the effects of the 
$\delchimax=6$ cut~\cite{sieve}, is given in the row labeled 
${\cal R}\times\chi^2_{\rm min}$/d.f. The errors in the fitted parameters are 
multiplied by the appropriate $r_{\chi2}$\cite{sieve}. }

\begin{tabular}[b]{|l||c||}     
	\multicolumn{1}{l}{Parameters}&\multicolumn{1}{c} {Values}\\
\hline
      $a_0$&$-5.381\times 10^{-2}\pm 2.17\times 10^{-3}$ \\ 
      $a_1$&$2.034\times 10^{-2}\pm 1.19\times 10^{-3}$\\ 
      $a_2$&$4.999\times 10^{-4}\pm 2.23\times 10^{-4}$\\
\hline
	$b_0$ &$9.955\times 10^{-3}\pm 3.09\times 10^{-4}$\\
      $b_1$&$3.810\times 10^{-3}\pm 1.73\times 10^{-4}$\\
      $b_2$&$9.923\times 10^{-4}\pm 2.85\times 10^{-5}$ \\ 
	\cline{1-2}
     	\hline
	\hline
	$\chi^2_{\rm min}$&165.99\\
	${\cal R}\times\chi^2_{\rm min}$&184.2\\ 
	d.f.&169\\
\hline
	${\cal R}\times\chi^2_{\rm min}$/d.f.&1.09\\
\hline
\end{tabular}
\end{center}
\end{table}
\def\arraystretch{1}  

The fit to the data on  $F_2^{\gamma p}(x,Q^2)$ was restricted to the region  $x\le x_P$.  In the absence of a global fit to the data in this region, they  simply extended their parameterization, piecewise, to the large-$x$ region, using the form
\ba
F_2^{\gamma p}(x,Q^2)&=&F_P\left(\frac{x}{ x_P}\right)^{\mu(Q^2)}\left(\frac{1-x}{1-x_P}\right)^3,\ \quad x_P<x\le 1, \label{Flarge}
\ea
where the piecewise extension on the r.h.s. of \eq{Flarge} is obviously continuous with the l.h.s. at $x=x_P$, independently of $\mu(Q^2)$.  The exponent $\mu(Q^2)$ is determined by requiring that the first derivatives  with respect to $x$ of the  function in Eqs.\ (\ref{Fofx}) and the function on the r.h.s. of (\ref{Flarge}) also match at $x=x_P$. These results  give the required  parameterization of $F_2^{\gamma p}(x,Q^2)$ over all $x$, required  in \eq{FF} to evaluate ${\cal F}(x,Q^2)$, and eventually, $\hat \calF (v,Q^2)$ of  \eq{hats}, needed to evaluate the exact solution for $\hat G(v,Q^2)$ in \eq{Gofv}, as well as calculating the Laplace transform $f(s,Q^2)$ used in  the numerical solution of \eq{g_solution}, which was calculated using the numerical algorithm of Appendix \ref {subsection:Laplacetransform}.

\begin{acknowledgments}

{\em Acknowledgments:} The author would like to thank Prof. L. Durand and Prof. Douglas W. McKay for their contributions to portions of this work, and also to thank the Aspen Center for Physics for its hospitality during the time parts of this work were done.  

\end{acknowledgments}

\bibliography{gluonsPRD.bib}

\end{document}